\documentclass[11pt,twoside]{article}
\usepackage{asp2004}
\usepackage{epsf}
\usepackage{psfig}
\begin{document}

\title{High-velocity \ion{O}{VI} in and near the Milky Way}

\author{Bart P. Wakker}
\affil{University of Wisconsin, Department of Astronomy, \\
475 N. Charter St., Madison, WI 53706, USA}
\author{Blair D. Savage}
\affil{University of Wisconsin, Department of Astronomy, \\
475 N. Charter St., Madison, WI 53706, USA}
\author{Kenneth R. Sembach}
\affil{Space Telescope Science Institute, \\
3700 San Martin Drive, Baltimore, MD 21218, USA}
\author{Philipp Richter}
\affil{Institut f\"ur Astrophysik, \\
Auf dem H\"ugel 71, 53121 Bonn, Germany}
\author{Andrew J. Fox}
\affil{University of Wisconsin, Department of Astronomy, \\
475 N. Charter St., Madison, WI 53706, USA}

\begin{abstract}
The Far Ultraviolet Spectroscopic Explorer (FUSE) has observed over three
hundred fifty sightlines to extragalactic targets. About one hundred fifty of
these are of sufficient quality to measure \ion{O}{VI} absorption in and near
the Milky Way. High-velocity \ion{O}{VI} absorption is detected in about 80\% of
the sightlines, with the detection rate going up for sightlines with higher
signal-to-noise ratios. \ion{O}{VI} is almost always seen in directions with
previously known \ion{H}{I} HVCs, including HVCs complex~C, complex~A,
complex~WD, the Magellanic Stream, and the Outer Arm. Studies of several
sightlines through complex~C suggest that the \ion{O}{VI} absorption is produced
in conductive interfaces between the cool HVC and a hotter surrounding medium,
most likely a corona around the Milky Way. The \ion{O}{VI} detections associated
with the Magellanic Stream imply that this hot corona has a radius of at least
50~kpc. About half of the detections of high-velocity \ion{O}{VI} are in
directions where no high-velocity \ion{H}{I} was previously known. Some of these
are probably associated with the Magellanic Stream, others may represent Local
Group gas. Still others often show up as a distinct wing on the low-velocity
absorption; these may either represent an outflow from the Milky Way associated
with the Galactic Fountain, they may be extragalactic HVCs, or they may be
tracing a wind from the Galactic Center. Distance limits to the high-velocity
\ion{O}{VI} are scarce, but at least one of the clouds appear to be more distant
than 4~kpc.
\end{abstract}

\section{Observations}
The Far Ultraviolet Spectroscopic Explorer (FUSE) satellite has now measured
\ion{O}{VI} absorption in and near the Milky Way in hundred forty five
sightlines to extragalactic objects. The \ion{O}{VI} $\lambda\lambda$1031.926,
1037.617 doublet are the best lines to use for kinematical investigations of hot
($T$$>$10$^5$~K) gas. Oxygen is the most abundant element heavier than helium,
and the \ion{O}{VI} lines have large oscillator strengths ($f_{1031}$=0.133,
$f_{1037}$=0.067, Morton 1991). FUSE was launched in 1999, and contains four
co-aligned spectrographs covering the 905--1187~\AA\ spectral region. The optics
in the two channels (side 1 and side 2) that record data above 1000~\AA\ are
coated with aluminum and lithium-fluoride (LiF), while the two channels
registering the flux below 1000~\AA\ are coated in silicon-carbide (SiC). Each
LiF and SiC channel is further split into two segments, corresponding to two
separate detectors (A and B). The \ion{O}{VI} lines are covered by four
channels, but a usefully high S/N ratio is only obtained in the LiF1A and LiF2B
segments. Spectra are sampled on 0.0068~\AA\ ($\sim$2~km\,s$^{-1}$) pixels, but
the resolution is about 15000 (20~km\,s$^{-1}$). FUSE data therefore always need
to be heavily rebinned. The S/N ratio obtained in the LiF1A segment is usually a
factor $\sim$1.5 higher than that obtained in the LiF2B segment.
\par As of May 2004, 345 distant extra-galactic objects had been observed by
FUSE. These include quasars, BL\,Lac objects, Seyfert galaxies, large
\ion{H}{II} regions in external galaxies, and starburst galaxies. This count
excludes stars in the Magellanic Clouds, M\,31 and M\,33. The signal-to-noise
(S/N) ratio of the observations ranges from 0 to a maximum of about 35. These
S/N ratios are measured per 20~km\,s$^{-1}$\ resolution element in the combined
LiF1A+LiF2B data. There are 13 objects for which an S/N ratio $>$20 near
\ion{O}{VI} has been achieved, 40 objects with S/N=10 to 20, 35 with S/N=6 to
10, and 57 with S/N=3 to 6. This gives a total of 145 objects for which Galactic
\ion{O}{VI} absorption can be measured. For the remaining 210 extragalactic
targets either the continuum near \ion{O}{VI} is too complicated, or the S/N
ratio is too low near \ion{O}{VI} (though it is sometimes higher elsewhere in
the spectrum). An earlier sample, consisting of 100 sightlines was described by
Wakker et al.\ (2003).
\par In almost all sightlines \ion{O}{VI} absorption is seen near velocities
relative to the LSR of $\vert$$v_{\rm LSR}$$\vert$$<$100~km\,s$^{-1}$. Just a
few (low S/N) sightlines yield an upper limit (see Savage et al.\ 2003). In
about 80\% of the sightlines \ion{O}{VI} is also seen at velocities
$\vert$$v_{\rm LSR}$$\vert$ larger than 100~km\,s$^{-1}$\ (see Sembach et al.\
2003). Figure~1 shows a representative set of spectra containing high-velocity
\ion{O}{VI}, while Fig.~2 shows both the \ion{H}{I} and the \ion{O}{VI}
high-velocity sky.

\begin{figure}[!th]
\plotfiddle{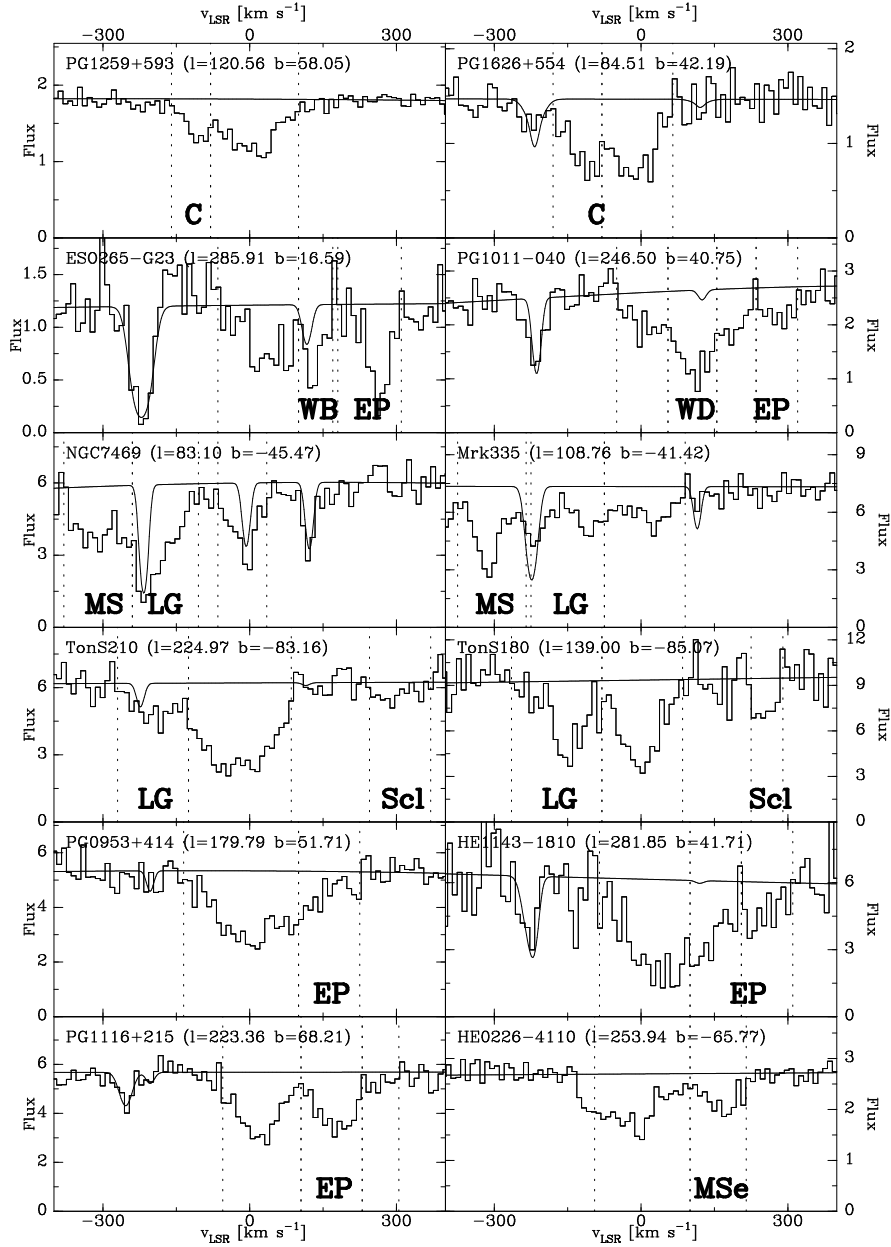}{6.4in}{0}{75}{75}{-220}{-60}
\caption{Sample of 12 sightlines showing the different kinds of high-velocity
\ion{O}{VI} detected by FUSE. The high-velocity components are identified by the
labels below the plots: C=complex~C, WB=complex~WB, WD=complex~WD, MS=Magellanic
Stream, MSe=MS extension, LG=possible Local Group, EP=extreme-positive velocity,
Scl=Sculptor group. The continua include a model for the H$_2$ absorption lines
at 1031.191 and 1032.356~\AA, and in one case (NGC\,7469) HD at 1031.912~\AA.}
\end{figure}

\section{High-velocity \ion{O}{VI} with high-velocity \ion{H}{I}}
\par In most cases where the sightlines intersect an HVC seen in 21-cm
\ion{H}{I} emission, a distinct \ion{O}{VI} absorption component is seen at the
same velocities. This is the case for twelve sightlines through HVC complex~C,
two through complex~WD, and seven through the Magellanic Stream. Only complex~A
is not always detected in both \ion{H}{I} and \ion{O}{VI}: just three (weak
\ion{O}{VI}) complex~A components are found in nine sightlines.

\begin{figure}[!th]
\plotfiddle{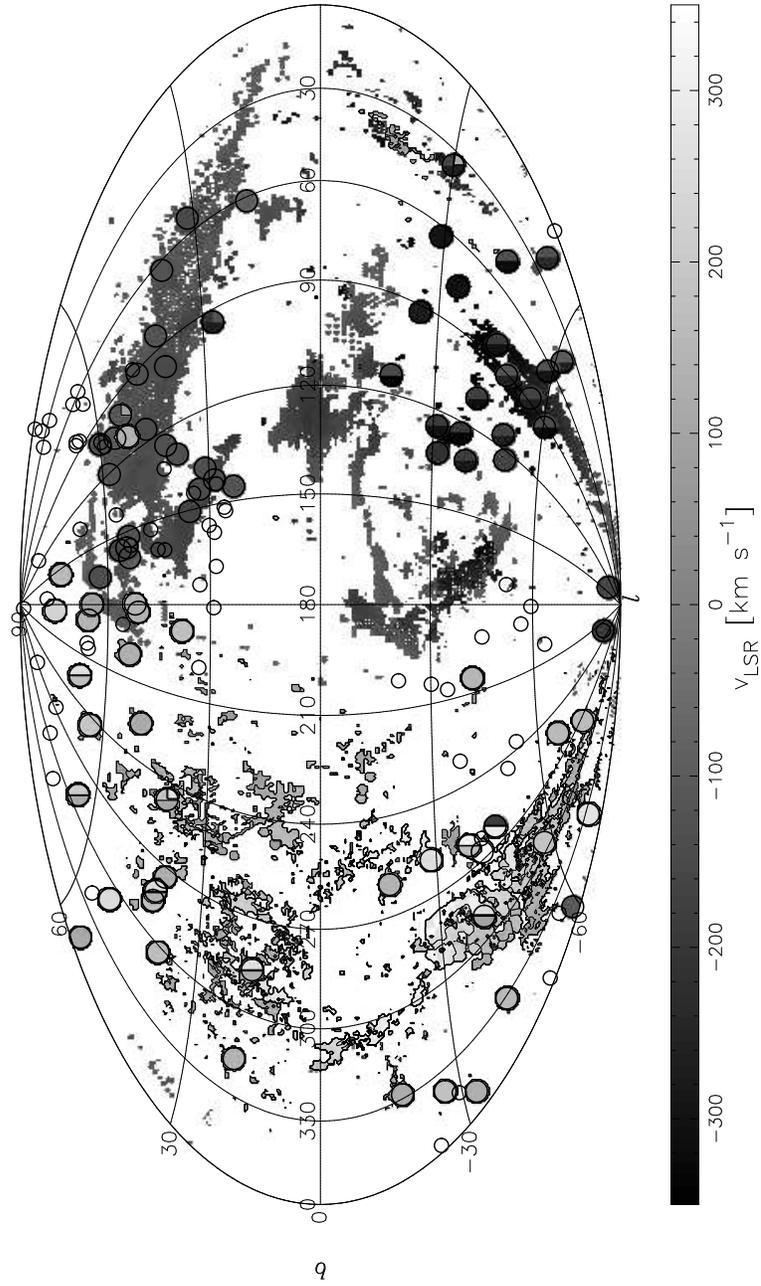}{6.9in}{0}{70}{70}{-210}{-55}
\caption{Comparison of high-velocity \ion{H}{I} and \ion{O}{VI}. The continuous
grey scale shows \ion{H}{I} LSR velocities from the Hulsbosch \& Wakker (1988)
survey. Filled circles indicate the sightlines with detected high-velocity
\ion{O}{VI}. Smaller open circles are for sightlines without high-velocity
\ion{O}{VI}.}
\end{figure}

\par Fox et al.\ (2004, and elsewhere in this proceedings) analyzed several of
the better complex~C sightlines, and compared \ion{O}{VI}, \ion{C}{IV},
\ion{N}{V}, and \ion{H}{I} with model predictions for the ionic ratios and
kinematics. They concluded that the \ion{O}{VI} most likely originates in a
conductive interface (or possibly in a turbulent mixing layer) between the cool
gas seen in 21-cm emission and a hotter (10$^6$~K or more) surrounding medium.
This result thus provides strong evidence for the existence of a hot,
highly-extended gaseous corona. Since the distance of complex~C is still unknown
(though it is $>$6~kpc, and is thought to be 10--20~kpc, see Wakker 2001), the
size of this corona is not constrained by the complex~C results. The fact that
\ion{O}{VI} is rarely seen associated with complex~A might imply that the hot
gas lies at $z$$<$5~kpc (for the $z$-height of complex~A, see van Woerden et
al.\ 1999). However, the fact that in directions through the Magellanic Stream
high-velocity \ion{O}{VI} is always seen in association with Stream \ion{H}{I}
would argue that the hot corona is larger than 50~kpc in diameter. Of course, it
is quite possible (likely even) that the hot corona is patchy.

\section{High-velocity \ion{O}{VI} without high-velocity \ion{H}{I}}
\par One of the more interesting discoveries coming out of the FUSE survey of
Galactic \ion{O}{VI} were the many absorption components in directions without
high-velocity \ion{H}{I} seen in 21-cm emission. Fifteen of these are probably
associated with the Magellanic Stream, as they lie within about 5 degrees from
the Stream as seen in \ion{H}{I}. Five positive-velocity wings are found at
galactic longitudes within 30\deg\ of the Galactic Center, and these could
possibly originate in a Galactic Wind. The more mysterious discoveries are the
18 detections with velocities of $\sim$$-$150~km\,s$^{-1}$\ at galactic
longitude $l$=30\deg\ to 140\deg, and galactic latitude $b$$<$$-$10\deg, and the
18 detections with velocities between +100 and +250~km\,s$^{-1}$\ at
$l$=175\deg\ to 330\deg, $b$$>$38\deg, often in the form of a ``wing'' to the
low-velocity absorption.
\par The detections at southern latitudes with velocities of about
$-$150~km\,s$^{-1}$\ lie in the same part of the sky as the Magellanic Stream.
But in 21-cm emission the Stream has velocities of $-$300 to
$-$450~km\,s$^{-1}$, and an \ion{O}{VI} absorption component is seen at those
velocities. (see Fig.~2). The tidal models do not predict any gas at lower
velocities in this part of the sky (e.g.\ Gardiner \& Noguchi 1996). On the
other hand, several Local Group galaxies with negative velocities are located
here, including M\,31, M\,33, and several dwarf irregulars (NGC\,6822, DDO\,210,
WLM, Pegasus, IC\,10), which have velocities ranging from $-$53 to
$-$237~km\,s$^{-1}$. Sembach et al.\ (2003) thus suggested that the
high-velocity \ion{O}{VI} might be associated with Local Group gas. However,
more analysis of the physical conditions in the \ion{O}{VI} gas is going to be
necessary before this conclusion can be made firm.

\begin{figure}[!th]
\plotfiddle{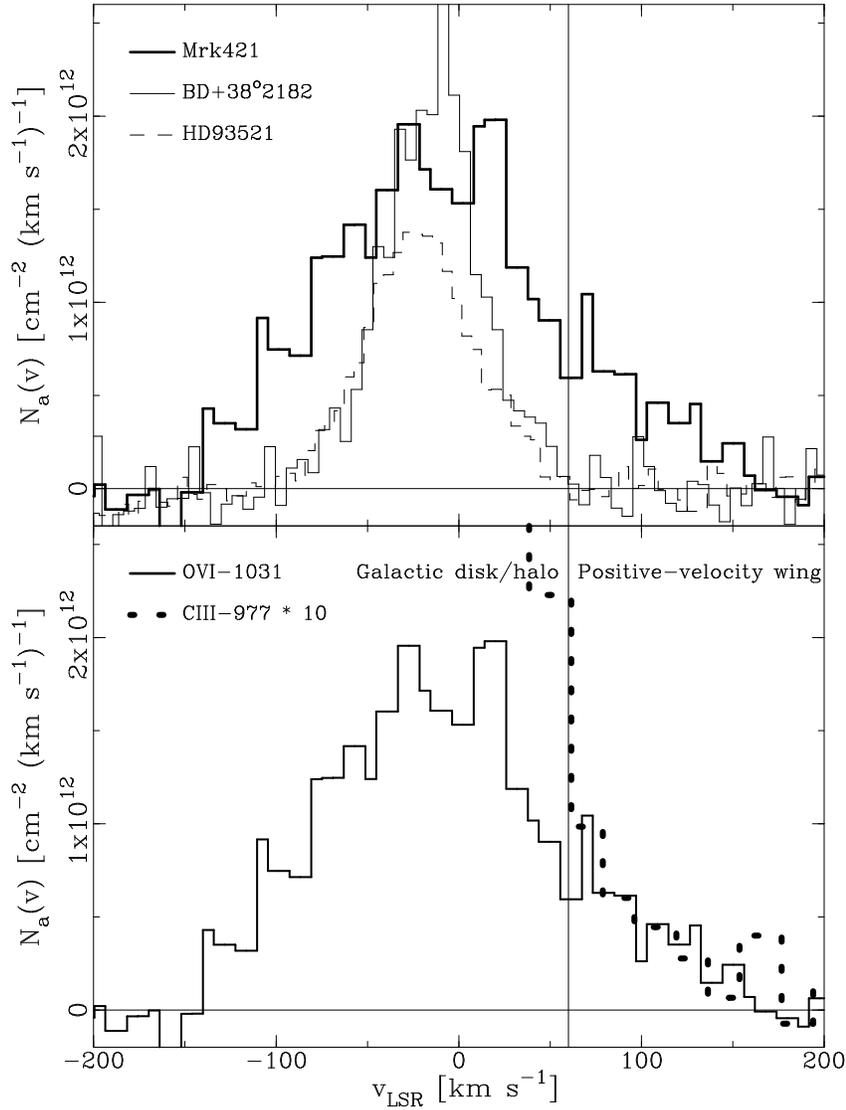}{6.3in}{0}{70}{70}{-190}{-50}
\caption{Apparent column density profiles as a function of LSR velocity for the
\ion{O}{VI} $\lambda$1031.926 line for Mrk\,421, BD+38\,2182 and HD\,93521. The
Mrk\,421 profile has been binned to 5 pixels (10~km\,s$^{-1}$), while the
profiles for the stars were binned to 2 pixels (4~km\,s$^{-1}$). Note the large
breadth of the \ion{O}{VI} profile for the extragalactic line of sight compared
to the halo star lines of sight. The lower panels compares the \ion{O}{VI} and
\ion{C}{III} absorption for Mrk\,421. The \ion{C}{III} profile has been scaled
by a factor 10. The vertical solid line at 60~km\,s$^{-1}$\ is where the
\ion{C}{III} transitions from very strong to weak.}
\end{figure}

\par The high-positive-velocity \ion{O}{VI} lies on the opposite side of the
sky. Again, gas in the Magellanic Stream generally would have high positive
velocities in this part of the sky, but the Stream is not seen in 21-cm emission
in the directions of the background targets, nor are such velocities expected
from the tidal model. There seem to be two explanations: the
high-positive-velocity \ion{O}{VI} may represent the outflowing Galactic
Fountain, or it may represent samples of distant gas clouds that even might be
the counterpart of the negative-velocity Local Group gas seen in the southern
sky. Again, more work is needed (and in progress) before these absorptions can
be given a definitive interpretation.

\section{The distance of the \ion{O}{VI} HVCs}
\par The distance to the gas producing the high-velocity \ion{O}{VI} absorption
is still very uncertain. For cases where \ion{H}{I} and \ion{O}{VI} are seen at
the same velocity it is possible to constrain the distance in the usual manner
(using detections and non-detections against stellar targets with known
distances, see Wakker 2001). For the \ion{O}{VI} associated with the Magellanic
Stream a distance of 50 to 150~kpc is implied by the tidal models of the Stream.
For the possible Local Group gas, and for the positive-velocity wings, however,
no such secondary information is available. For these absorbers there are two
possible ways to constrain the distance. First, if many other ions are seen to
be associated with the \ion{O}{VI} absorption, one can analyze the physical
conditions in the cloud, combining the ionic column densities with
photoionization models, such as those provided by CLOUDY (Ferland et al.\ 1998);
this gives constraints on the cloud's size. Second, one can search for the
presence or absence of high-velocity \ion{O}{VI} absorption in the spectra of
distant stars. So far, only one such comparison has been published (Savage et
al.\ 2004), shown in Fig.~3. The absorption toward the QSO Mrk\,421 shows a
clear positive-velocity wing, extending from +65 to +165~km\,s$^{-1}$. No such
wing is seen in the spectrum of the star BD+38\,2182, just 4\deg\ away, at a
distance of 4.0~kpc. This clearly indicates that the high positive velocity
\ion{O}{VI} gas is distant.

\end{document}